IAC-22-B1.3.10

# Infrared Remote Sensing Using Low Noise Avalanche Photodiode Detector

**Joice Mathew[a]\*, James Gilbert [a], Robert Sharp[a], Alexey Grigoriev[a], Nicolas Cardena[b], Marta Yebra[b,c]**

[a] *Advanced Instrumentation and Technology Centre, Research School of Astronomy and Astrophysics, Australian National University, Canberra, ACT 2611, Australia*, joice.mathew@anu.edu.au, rob.sharp@anu.edu.au, alexey.grigoriev@anu.edu.au

[b] *Fenner School of Environment and Society, The Australian National University, Acton, Australian Capital Territory, Australia,* nicolas.younes@anu.edu.au, marta.yebra@anu.edu.au

[c] *School of Engineering, The Australian National University, Acton, Australian Capital Territory, Australia*

\* Corresponding author

### Abstract

For a remote sensing optical payload to achieve a Ground Sampling Distance of ~ 10-30 m, a critical problem is platform-induced motion blur. While forward motion compensation can reduce this transit speed, it comes at the expense of a more challenging satellite attitude control system and induces a variable observation/illumination angle. This relative motion can be frozen out by simply reading the sensor system at a frame rate that matches the ground resolution element's pixel crossing time. To achieve high resolution using this Time-Delay Integration (TDI)-like approach requires high speed and hence near "zero" readout noise detector arrays to avoid swamping the observed signal. This requires associated control electronics for fast frame readout and direct interface with smart- Artificial Intelligence (AI) onboard processing. With this technique, the platform freezes out its movement concerning the ground, reducing the demands placed on the attitude control systems, which can otherwise be difficult to implement on a small satellite platform. Here we report the Australian National University's OzFuel mission which applies this technical solution to deliver high ground resolution via high frame rate imaging. OzFuel is built around the Leonardo SAPHIRA Mercury Cadmium Telluride linear mode electron avalanche photodiode (LMeAPD) detector and the in-house developed Rosella electronics control system. The mission will deliver an integrated sensor system in a suite of Short-Wave Infrared (SWIR) passbands dedicated to monitoring the flammability of Eucalypt trees. The OzFuel mission concept focuses on the application of SWIR remote sensing data to deliver a strategic evaluation of fuel loads and moisture content in the bushfire-prone Australian environment. Here we will provide an overview of the SAPHIRA detector and its application in SWIR remote sensing along with an overview of the OzFuel mission.

**Acronyms/Abbreviations**

| | |
|---|---|
| ADC | Analog-to-Digital Converter |
| AI | Artificial Intelligence |
| ANU | Australian National University |
| CMOS | Complementary Metal-Oxide-Semiconductor |
| COTS | Commercial Off the Shelf |
| EO | Earth Observation |
| FOV | Field of View |
| FMC | Fuel Moisture Content |
| FPGA | Field Programmable Gate Array |
| GSD | Ground Sampling Distance |
| LEO | Low Earth Orbit |
| LMeAPD | Linear-Mode electron Avalanche Photodiode |
| NIR | Near-Infrared |
| NSTF | National Space Test Facility |
| ROIC | Readout Integrated Circuit |
| SNR | Signal to Noise Ratio |
| VNIR | Visible Near-infrared |
| SWIR | Short Wave Infrared |
| TDI | Time Delay Integration |

## 1. Introduction

Despite recent technological advances and the launch of several Shortwave Infrared (SWIR) capable satellites, the micro to SmallSat market has yet to fully embrace SWIR imaging capabilities, leaving a gap in this important spectral region. In addition to its distinctive spectral properties, SWIR remote sensing can provide information that is not possible with visible, near-infrared or thermal imagery [1]. Collecting satellite imagery in SWIR wavelengths has unique benefits, including improved atmospheric transparency and material identification. The SWIR remote sensing can provide critical diagnostic data for agriculture and moisture content monitoring on a continental scale [2, 3]. It has also applications to mineralogical exploration to provide critical, analytical information that provides a rapid exploration of large remote areas [4]. SWIR remote sensing data can provide crucial information on fuel flammability and heat sources across continents which is critical for bushfire behaviour and risk monitoring [5] and fire detection [6].

To get sensitive observations, cooled detector systems are often needed for SWIR remote sensing. Here SWIR refers to electromagnetic radiation with a wavelength of approximately 1100 to 2500 nanometers (nm) [7]. Cooling and waste heat rejection requirements have historically limited SWIR missions to larger platforms. These platforms, though useful for certain applications, are not well equipped to test new mission concepts or rapidly respond to emerging needs. It is highly desirable to have a Ground Sampling Distance (GSD) down to 10 m for a SWIR Earth observation payload in low Earth orbit (LEO) and this would bring a significant shift in the value proposition for remote sensing data.

One of the challenging issues in achieving 10-20 m GSD is the platform-induced motion blur. To avoid motion blur and achieve image stability at LEO platforms, active satellite tracking and pointing control is required during observation. This would demand stringent requirements, especially on active pointing and tracking of the satellite platform and could potentially increase the complexity and cost of the mission. Furthermore, the constantly shifting illumination attitude during operations presents calibration and parallax issues that compromise the quality and integrity of data.

An alternative to complex tracking systems is to use the time-delay imaging (TDI) technique to "freeze-out" platform orbital motion blur. The platform-induced blur can be frozen out by simply reading the sensor system at a frame rate that matches the ground resolution element's pixel crossing time. However, framerates of up to 1 kHz are required to achieve <10 m ground resolution from LEO. While such framerates are achievable at visible near-infrared (VNIR) and SWIR wavelengths, the challenge is to reduce the high readout noise associated with conventional CMOS electronics.

The SAPHIRA infrared (0.8-2.5 µm) electron avalanche photodiode (eAPD) array, from the UK manufacturer Leonardo MW (LMW), uses a high avalanche gain stage that essentially improves the electronic noise properties of the system [8-10]. This makes it possible to achieve <10 m ground resolution at LEO operations from a small form factor package requiring only basic pointing stability control and hence relaxing the power and control requirements.

The OzFuel mission applies this technical solution to deliver high ground resolution via high frame rate imaging. OzFuel is built around the Leonardo SAPHIRA avalanche photodiode detector and the in-house developed Rosella electronics control system [10]. The mission will deliver an integrated sensor system in a suite of SWIR passbands dedicated to monitoring the flammability of Eucalypt trees.

## 2. TDI Imaging for Earth Observation

Time Delay Integration (TDI) or "drift scan" imaging is an electronic scan technique, based on the concept of the accumulation of cumulative exposures of the same object as it is moving across the FOV. While the CMOS readout architecture of the Leonardo eAPD precludes conventional TDI imaging operations, the low effective readout noise of the eAPD (when operated with avalanche gain) means that quasi-TDI observations can be implemented effectively by reading the full array at the pixel crossing time and shifting and stacking images. Without avalanche gain, conventional CMOS readout noise levels result in a highly read noise-limited final observation, with little to no sensitivity. Also, the conventional CMOS array operations would need the incoming field to be stabilized onto the focal plane array for the full integration time. This requires a complex control system and also introduces additional demands on image stability and sensitivity variation across the tracked field. One of the drawbacks of TDI imaging is the requisite for considerable parallel onboard data processing to reduce the value-added frame data volume, but this can be addressed by state-of-the-art image compression and Artificial Intelligence (AI) based processing systems.

SAPHIRA detectors can operate in TDI mode and they are already used in ground-based astronomical instruments for fast wavefront sensing and fringe tracking applications, but their utility in space applications is yet to be exploited. Specifically, their suitability for high-speed and/or photon-starved applications has profound implications for space-based IR instruments, including TDI-like surveys of the sky from a low earth orbit platform. A use case of the TDI capability of SAPHIRA for infrared astronomical observation from space is the Emu mission by the Australian National University (ANU) [7].

### 3. SAPHIRA: A Low Noise Avalanche Photodiode Detector

SAPHIRA Avalanche Photo Diode arrays (Fig. 1), from Leonardo have already transformed ground-based astronomy due to their 'noise-free' imaging performance and ability to read out images at high frame rates [11]. Indeed, the early development of SAPHIRA devices was funded for use in high-speed adaptive optics and interferometry instruments for ground-based astronomy [12, 13].

The key advance in these sensors is a novel electron-avalanche gain layer in the device that boosts signals by several hundred times before they arrive at the sensor pixel [10]. This overcomes what has been a fundamental limitation of traditional infrared sensors, namely that the very act of accessing pixel values introduces significant and unavoidable readout noise due to physical processes in the pixel electronics. The equivalent level of this noise concerning the incoming light is typically many tens-of-photons even in the very best devices, setting a fundamental limit to the object brightness that can be detected, and historically driving up exposure times to compensate.

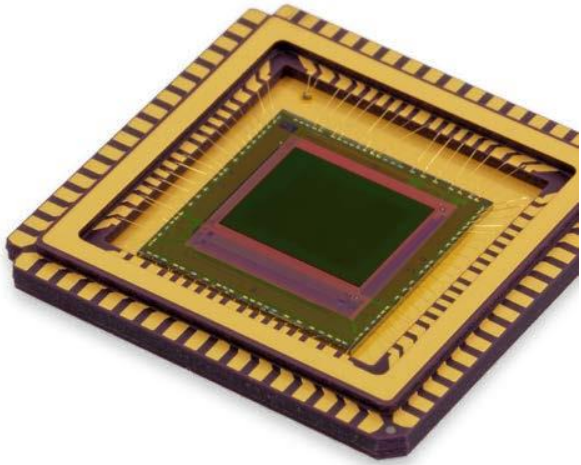

Fig. 1. The SAPHIRA is a 320×256 pixel linear-mode avalanche photodiode array. Image courtesy Leonardo MW [11].

The SAPHIRA's ability to amplify signals before the injection of readout noise effectively reduces the readout noise contribution to zero [14], providing an unprecedented level of sensitivity. Furthermore, this extra sensitivity can be used to reduce exposure times and therefore greatly increase the frame rate of the system for any given signal level; the SAPHIRA itself has a theoretical maximum readout rate of 3,900 frames per second [8]. This opens a swathe of previously impossible use cases, and while these are being explored for a growing number of applications on Earth, ANU is leading the way in demonstrating their applications in space [8-10] and the OzFuel mission [15].

The SAPHIRA detector has 320×256 pixels with a 24 μm pitch [10]. It is a HgCdTe sensor, sensitive to a wavelength range of 0.8–2.5 μm. It has a CMOS readout integrated circuit (ROIC) which provides several readouts and reset modes [10]. The SAPHIRA adds an eAPD region below a traditional HgCdTe absorption layer, as part of a hybridized assembly with a CMOS readout integrated circuit with 32 parallel readout channels, each supporting a pixel rate of up to 10 MHz [8]. A fully integrated SAPHIRA focal plane built at ANU is shown in Fig. 2.

Current development efforts at Leonardo aim to further reduce dark current while increasing array size to 1k×1k and beyond, paving the way for the next generation of space-borne infrared instruments.

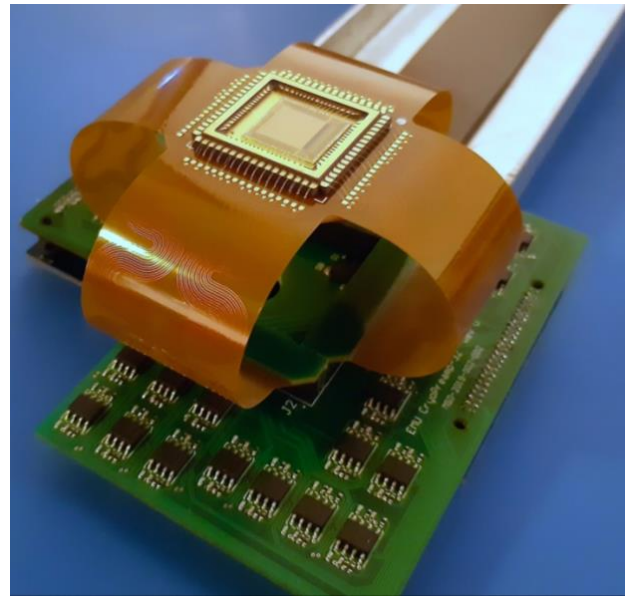

Fig. 2. The fully integrated SAPHIRA with flex cable and cryogenic preamplifiers.

### 4. Rosella Readout Electronics

Rosella is a modular and compact detector controller for space applications developed by ANU. This high-performance Field-Programmable Gate Array (FPGA) based readout system can be configured to interface with wide range of visible and infrared CMOS detectors including SAPHIRA and Teledyne HxRG family of infrared arrays.

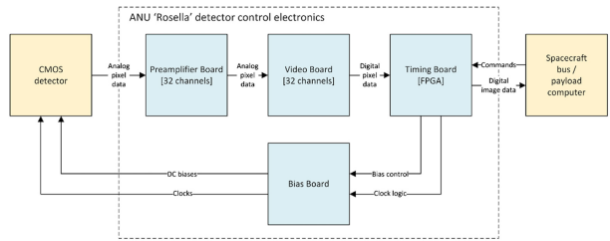

Fig. 3. The architecture overview of Rosella

Rosella electronics architecture includes a preamplifier board, bias board, video board, and an FPGA-based timing board (Fig. 3). The preamplifier board reduces the effect of electrical noise on detector output signals by increasing the amplitude of these signals early in the signal chain. The bias board is responsible for generating stable and accurate DC voltages for the SAPHIRA detector and variable gain bias.

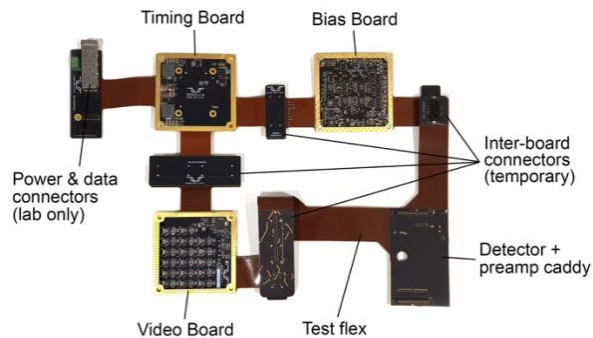

Fig. 4. Rosella 'FlatSat' engineering model.

The video board has 32 Analogue to Digital Converter (ADC) channels for digitizing pixel stream coming from the detector [8]. All these boards together deliver a low readout noise system, which is critical for SAPHIRA-like detectors. The timing board is responsible to manage the entire system, including clock pattern generation, bias configurations, ADC triggering, image processing, and communication with an external payload computer via a standard protocol. A 'FlatSat' model of Rosella is shown in Fig. 4.

For the Emu astronomy space mission, Rosella will be used to acquire frames at 50 Hz [8]. However, Rosella has been designed with high-resolution Earth Observation missions in mind, which can support frame rates up to 1 kHz. A breadboard prototype of Rosella has been developed and successfully demonstrated on-sky at the ANU 2.3 m telescope for 'lucky imaging' application [10].

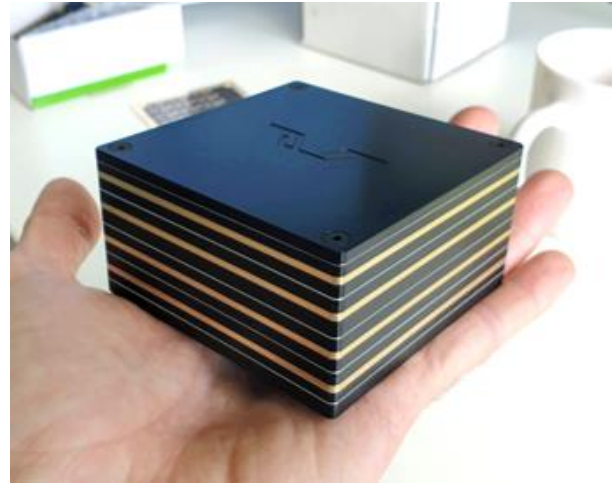

Fig. 5. Rosella 0.5 U enclosure mockup showing interleaved PCBs and aluminium enclosure walls

Rosella's final version will occupy a volume of ~0.5 U, comprising a connector-less printed circuit board (PCB) assembly based on rigid-flex technology [8]. Rigid PCB sections have an outer thermal conduction region that interfaces with an aluminum wall, forming a contiguous and enclosed board stack by folding the flex circuit sections. The enclosure also provides the right tightness for payloads sensitive to infrared emission (thermal "glow"). A mock-up of the PCB and enclosure assembly is shown in Fig. 5.

5. **OzFuel Mission Overview**

The OzFuel mission is the first step in the creation of an early warning Earth observation (EO) system, with optimum revisit, and high spatial and spectral resolutions to assist the government and communities in enhancing bushfire situational awareness and preparedness. [15]. OzFuel-1 is the first version of OzFuel mission and as pathfinder project, it will be operating at four specific narrow SWIR bands sensible to changes in the key fuel properties that determine the eucalypt's flammability and fire risk (e.g. dry matter, water and oil content).

OzFuel phase-1 shares many operational design characteristics with the CHICO hyperspectral visible light satellite also being developed at ANU. A road map of the ANU remote sensing satellite development program is shown in Fig. 6. Future phases of the OzFuel mission will use SWIR hyperspectral sensing and in the long term, a constellation of hyperspectral satellites.

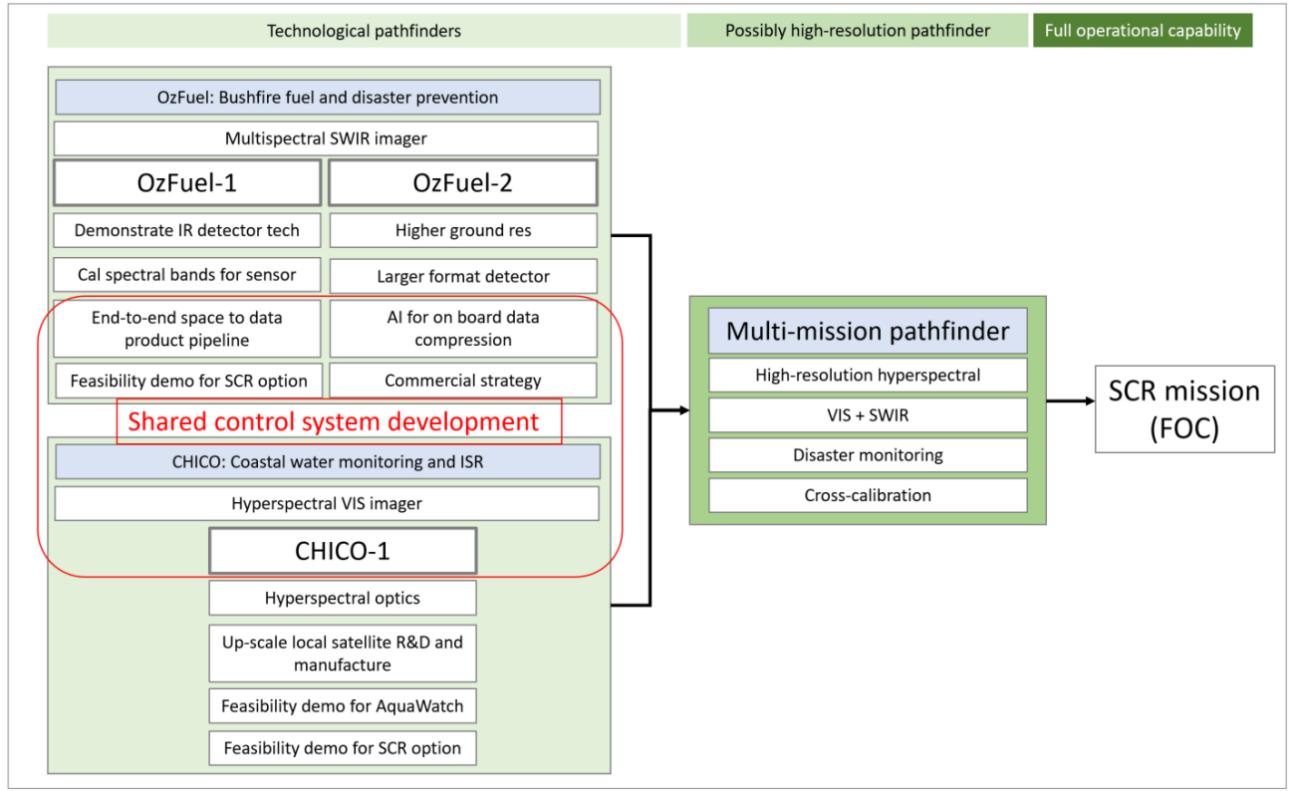

Fig. 6. Road map of ANU remote sensing satellite [15]

## 6. Mission System Requirements

The OzFuel mission aims to monitor the flammability of eucalypt forests via satellite remote sensing to deliver whole-of-continent bushfire fuel load data. the optimum spatial, temporal, and spectral resolution. Forest fuel flammability in eucalypt forests is not expected to change suddenly but gradually over a period of days or weeks. For a pathfinder mission, a temporal resolution of 6-8 days is acceptable. Repeat coverage of the same spot by an imaging satellite should be higher than that of the Landsat sensors (i.e. every 16 days) and similar to that of the combined Sentinel 2A and 2B satellites (i.e. every three to five days) [15]. A constellation of two or more satellites can accomplish this.

As some flammability parameters such as fuel moisture content (FM) changes throughout the day [16, 17], it would be desirable to acquire data in the early afternoon, when vegetation is most stressed and can therefore be more easily ignited. Preferable time of observation for the mission is between 12h00 and 14h00. The schedule will vary depending on cloud cover, potential 'hot spots' in images, and other factors that might complicate image analysis.

A Ground Sampling Distance (GSD) between 20 to 60m is acceptable for the OzFuel pathfinder mission and a GSD of 50 m would provide enough information for fuel flammability monitoring [15]. A swath width between 100 km and 150 km, would be desirable [15]. Given the constraints of the currently available SAPHIRA sensor, a swath width of 16 km can be achieved with a GSD of 50 m per pixel. A higher swath can be achieved by using a larger SAPHIRA array (eg: 512 x 512 or 1K x 1K format sensors). To monitor changes in flammability, OzFuel targets specific wavelengths that provide information related to vegetation water content, dry matter content and other metabolites (oils, volatiles).

Ozfuel will have four spectral bands linked to those parameters (Table 2). These spectral bands will provide basic inputs for the proposed products, and it is crucial for 'pre-fire' management activities and the characterisation of fuel loads across Australia. As shown in Fig. 7 the selected spectral bands for OzFuel will allow good separability of leaves with varying moisture and dry matter content.

Table. 1. Summary of OzFuel mission requirement

| Characteristic | User requirement |
|---|---|
| **Revisit time** | 6-8 days |
| **Time of observation:** | Diurnal observations, preferably between 12h00 and 14h00 |
| **Ground sampling distance:** | 50m |
| **Swath width:** | At least 16 km |
| **Albedo** | 10-20% for eucalypt forests |
| **Spectral range:** | 1200 – 2300 nm |
| **Spectral band centre:** | 1205 nm |
|  | 1660 nm |
|  | 2100 nm |
|  | 2260 nm |
| **Number of Spectral bands:** | 3 – 4 bands |
| **Radiometric resolution:** | 12 to 16 bit |
| **Signal-to-Noise ratio:** | 100:1 or better |
| **Geographical coverage:** | Australian mainland and Tasmania |

A signal-to-noise ratio (SNR) of at least 100 will be achieved in each spectral band. A high radiometric resolution will be required for OzFuel to detect slight changes in the flammability of eucalypt forests. To ensure proper sampling of the dry and wet properties of the vegetation, between 12 and 16 bits of radiometric resolution are needed.

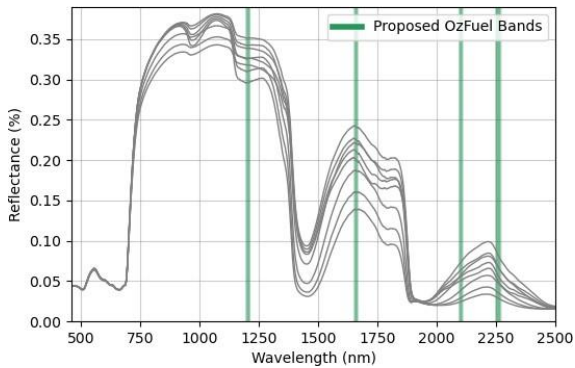

Fig. 7. The proposed four spectral bands of the OzFuel mission are shown in Green colour over the spectra of leaves with different water and dry matter content [15]. The different grey lines are spectra for leaves with different water and dry matter content, two key drivers of leaf flammability.

The mission will deliver flammability data globally during its mission lifetime. A summary of mission requirements is shown in Table. 1.

**7. Payload**

The OzFuel mission concept is based on the SAPHIRA SWIR detector as discussed in Section 3. A second-generation OzFuel satellite (OzFuel-2) would seek to deploy larger form factor arrays (1k and 2k devices are available in the same family). A large array footprint implies a wider swath width or a smaller GSD. Once the control electronics architecture and thermal management system have been verified in the less demanding OzFuel-1 missions, a larger SAPHIRA array can be deployed in the subsequent mission.

The payload for the pathfinder OzFuel mission will be a SWIR multispectral imager. The imaging optics will have ~85 mm square. It delivers sufficient ground resolution (<50 m at all SWIR wavelengths) while also providing adequate image sensitivity. Moreover, it conforms to the industry-standard form factor "1U" and can be used in a "3U" configuration without the need for high-risk fast optical systems. The optical system will be well matched to the performance specifications of a high-grade SWIR focal plane detector driven by a dedicated high-speed electronics package (like Rosella). The instrument will be designed such that there will be no significant added development cost to support a 512 x 512 pixel 24 μm eAPD array.

The OzFuel mission when operating in TDI mode avoids the need for costly (in terms of volume, weight, complexity and direct expense) attitude control systems that are challenging on a small satellite platform. The required high frame rate of the detector system places high demands on the associated control electronics. Commercial off-the-shelf (COTS) control solutions typically cannot deliver the low-level detector control necessary to provide low latency data digitization, particularly when interfacing with external on-board artificial intelligence processing systems or when real-time data manipulation is required. The mission will have the capability to operate both in snap-shot and rolling shutter (TDI) imaging mode.

As OzFuel will be operating in the SWIR spectral region the thermal emission from the low emissivity telescope optics is expected to be of limited concern at the operating temperature of 20-40 °C expected for a LEO smallSat in radiative equilibrium. However, high-performance SWIR detectors capable of operating out to 2.5 μm required cooling to temperatures of ~100-200 K to suppress excessive dark current to usable levels. Additionally, the detector enclosure is a unit emissivity surface that must be controlled to prevent flooding the sensor with thermal background photons. OzFuel will adopt the thermal control system developed by Melbourne Space Laboratory (MSL), which employs the TheMIS thermal control module coupled with the Thales LSF9987 cryogenic cooler. TheMIS was developed for MSL's SpIRIT mission [18].

Canberra-based Skykraft will provide the satellite bus for the mission. Leveraging Skykraft technology will reduce non-recurring engineering costs to develop a custom platform for the OzFuel mission. The first OzFuel satellite is envisioned as a microsatellite (<100 kg) operating a 4-band SWIR sensor in a LEO. The orbit would be selected to enable the demonstrator to image calibration sites in Australia more frequently. An orbit for a constellation would be chosen to provide a 3-5 day revisit rate and possibly provide coincident observations with highly-calibrated optical missions such as Landsat 8 and Sentinel 2.

The payload will be assembled and calibrated in a class 10000 clean room at the ANU National Space Test Facility (NSTF). A geometric, radiometric and ground calibration will be performed to ensure the payload system performance. Space qualification (such as vibration, thermo-vac and EMI/EMC test) of each subsystem and the whole satellite will be performed at the NSTF. The OzFuel payload will have the capability to perform periodic instrument on-orbit geometric, radiometric, and spectral calibration.

## 8. Summary and Future Work

SAPHIRA LMeAPD detector array from Leonardo UK has profound implications for space-based IR instruments, especially when there is a demand for high-speed and/or photon-starved observation from space. These detectors along with ANU-developed Rosella control electronics can support TDI-like imaging to significantly reduce the pointing and tracking demands/complexities associated with small-satellite platforms. SAPHIRA and Rosella can enable crucial SWIR remote sensing observation from space for bushfire risk monitoring. ANU is leveraging these in-house technological capabilities to lead the development of a sovereign EO bushfire monitoring system.

OzFuel mission aims to monitor whole-of-continent eucalypt fuel flammability via satellite remote sensing every six to eight days during the early hours of the afternoon when vegetation is most stressed and more easily ignites. Images would be taken at a spatial resolution of about 50 metres, which is adequate for bushfire management operations. Conceptualized as a pathfinder to a national environmental monitoring constellation, the mission will provide critical bushfire EO data to support the government, frontline organisations and communities for enhanced bushfire situational awareness and preparedness. OzFuel is being developed in parallel with the CHICO mission, a dual-use hyperspectral imager for water quality monitoring. Currently, Phase A study of the mission is finished and prototyping of critical and high-risk sub-systems is underway.